\documentclass[12pt]{article}
\include{epsf}
\def\beq{\begin{eqnarray}}
\def\eeq{\end{eqnarray}}
\usepackage{amssymb}

\begin{document}

\title{On the dynamics of a self-gravitating medium with
random and non-random initial conditions}
\author{E. Aurell$^{1,2}$, D. Fanelli$^2$ \&  P. Muratore-Ginanneschi$^3$}
\maketitle

\begin{center}
\begin{tabular}{ll}
$^1$ & Dept. of Mathematics, Stockholm University,\\ 
& $\qquad$ SE-106 91 Stockholm, Sweden\\
$^2$ & Dept. of Numerical Analysis and Computer Science, 
        Stockholm University/KTH,\\
& $\qquad$ SE-100 44 Stockholm, Sweden\\
$^3$ & Niels Bohr Institute, Blegdamsvej 17, \\
& $\qquad$ DK-2000 Copenhagen, Denmark
\end{tabular}
\end{center}

\begin{abstract}

The dynamics of a one-dimensional self-gravitating medium, with 
initial density almost uniform is studied.
Numerical experiments are performed with
ordered and with Gaussian random initial conditions. 
The phase space portraits are shown to be qualitatively 
similar to shock waves, in particular with initial
conditions of Brownian type.
The PDF of the mass distribution is investigated.
\end{abstract}

\begin{flushleft}
Submitted to Physica D
\end{flushleft}

\section{Introduction}
\label{s:introduction}
In this paper we  study the dynamics
of a self-gravitating medium, the initial density of which is almost 
homogeneous, and of which the initial velocities of all fluid particles 
are small.
 
The study is performed in one spatial dimension where gravitational
force is a Lagrangian invariant. 
The equations of motion
for the density and velocity perturbations can therefore be written 
such that the net acceleration of fluid particles are Lagrangian
quasi-invariant, if the perturbations are represented as discrete mass 
points moving on a stationary and uniform background. This system of 
particles is then brought forward in discrete time steps from one collision
to the next, which, 
as shown already by Eldridge and Feix
\cite{EldridgeFeix},
permits the construction of an exact (up to round-off 
errors) and fast numerical integration scheme.

One reason to study the one-dimensional case is, that
we wish to study the dynamics starting from  
initial conditions
that are Gaussian random fields with power-law spectrum,
and the statistical properties of these
solutions. By analogy with recent work on the adhesion model
\cite{SheAurellFrisch,VergassolaDubrulleFrischNoullez,GurbatovSimdyankinAurellFrischToth},
we then expect that there is a wide range of
spatial scales in the solutions at late times.
To get good statistics, while still resolving fully
all details of the motion, a one-dimensional model
is more convenient than simulations in two or
three dimensions.

Our motivation to study gravitational dynamics with such random
initial conditions is that this is as a model of
structure-formation in the early universe, see 
\cite{Weinberg1972,Peebles1980,ZeldovichNovikov}. We will
return to a discussion of background 
material in cosmology in section~\ref{s:cosmological}. 
A second reason for a detailed study of one-dimensional case is then that,
in general position, gravitational collapse
accentuates asymmetries in the velocity and density fields. The first 
structures to form are blinis or pancakes, thin in one direction
and of large extent in the two others.
What we study can then be pictured as 
the one-dimensional dynamics of
very large blinis, oriented parallel to one another and colliding and
merging when moving under their mutual gravitational attraction.
For very long times the tree-dimensional nature of the motion is
no doubt important, but for some time after the first blinis form,
the one-dimensional approximation should be appropriate. This point
has recently been extensively discussed in 
\cite{VergassolaDubrulleFrischNoullez}, 
in the context of the adhesion approximation.
The one-dimensional model can hence perhaps also give a quantitatively
correct description of the clustering of mass as function of time
as long as we consider the largest structures at every moment of time.

The main new results of this paper are as follows:
we rederive the Rouet et al. 
mathematical model in a 
 way which appear transparent to us, with particular
emphasis on initially localized perturbations.
We discuss the differences between structures observed
on a uniform and homogeneous 
background, and those in a finite medium without a background.
As has been observed previously
\cite{Rouet1990,Rouet1991}, phase space portraits, starting from
ordered initial conditions, or random initial conditions, 
consist of smooth one-stream intervals (with ordered initial conditions), 
and short intervals with multi-stream solutions and high mass concentration.
In particular, with Gaussian initial conditions of strong spectral support
at low wave numbers, e.g. of Brownian type, we find
qualitative similarities to the mass distributions in
the adhesion model, i.e. ramps and mass concentrations of all
sizes at all scales. We try to quantify the mass distribution
by computing scaling exponents and mass histograms in octaves. 

Recent mathematical investigations
of finite self-gravitational systems, without
a background, are~\cite{Roytvarf} (ordered
initial conditions) and~\cite{BonvinMartinPiaseckiZotos}
(random initial conditions, but of another class than
we use), for a discussion of applications of this
system to structure formation in the universe,
see~\cite{GurevichZybin}. For Vlasov-Poisson equation in
one dimension (mutual repulsive interaction), 
see~\cite{MajdaMajdaZheng}.

The paper is organized as follows. 
In section~\ref{s:cosmological} we summarize standard material
in cosmology, and discuss references to recent observational data on 
the anisotropies of the cosmic microwave background radiation. 
In section~\ref{s:generalized-zeldovich} we 
derive the Rouet et al. solution of 1D self-gravitating
systems, and discuss initial conditions appropriate for our
system and admissible boundary conditions.
In section~\ref{s:numexp} we study the
dynamics starting from ordered and random initial conditions,
and in section~\ref{s:mass} we investigate properties of the
mass distribution.
In section~\ref{s:discussion} we sum up and discuss our results.

\section{The cosmological and observational setting}
\label{s:cosmological}

In our immediate neighborhood today, the universe is neither
homogeneous nor isotropic. Sources of electro-magnetic radiation in any 
frequency band are distributed in a markedly random and clustered manner 
over the sky.
On the other hand, at large scales the universe is generally 
taken to be homogeneous and isotropic. 
The hypothesis of an early almost homogeneous and
isotropic state of the universe rests on that it agrees with
the whole body of theory of the hot Big Bang, and 
with the observed 3K black-body background radiation. 
It is therefore natural and standard to assume that 
the structures observed today are due to instabilities in an
initially almost homogeneous self-gravitating 
medium \cite{Weinberg1972,Peebles1980,ZeldovichNovikov}.

The study of such instabilities has a long pre-history, 
going back already to Newton~\cite{Newton}.
The first quantitative investigation of the instability in a static medium 
with non-zero pressure was performed by Jeans~\cite{Jeans}, who derived his 
famous formula that perturbations of wave-length larger than  
$\lambda_J = v_s\sqrt{\pi/G\rho}$ are unstable, where $v_s$ is the sound 
speed, $G$ the gravitational constant and $\rho$ the density. 
Such perturbations hence grow exponentially in time (in the Jeans theory), 
while perturbations of smaller wave-length oscillate and do not grow.
The Jeans length can be related to a Jeans mass of 
$M_J = \frac{4\pi}{3}\rho \lambda_J^3$. In an almost homogeneous gas, 
regions of increased density with mass larger than the Jeans mass collapse 
gravitationally, while concentrations of smaller mass oscillate acoustically.

While Jeans' analysis immediately applies to 
gravitational collapse of a finite object, it is not
complete when referring to a medium of infinite extent.
Since a sufficiently large mass will always be unstable,
the unperturbed state assumed in the Jeans formula, an infinite 
self-gravitating medium with gravitational self-interaction and
constant density, cannot exist in classical physics.
On the other hand, in general relativity the Friedmann solutions to the Einstein
equations describe
unbounded universes with spatially
constant density. These solutions are given in terms of a 
cosmological length scale $a$, which changes with cosmological time $t$.
On sufficiently small length scales (much less than $a$), and on
sufficiently short time-scales (faster than $\frac{a}{\dot a}$) 
general relativity is well approximated by Newtonian gravity,
and Jeans' analysis of an infinite medium
is hence well-founded in this way.

The linear theory of small perturbations around the Friedmann
solutions in general relativity was 
developed by Lifshitz, and is described in detail in 
Weinberg \cite{Weinberg1972} and Zeldovich \& Novikov \cite{ZeldovichNovikov}.
If we assume a hydrodynamic description of the matter fields, the
perturbations can be classified as scalar, vector and tensor.
The last ones correspond to gravitational waves, which
have no counterpart in the classical theory,
and will be left aside in the following.
 
Although the relevant calculations are involved, to a considerable 
degree the results can simply be described by introducing the co-moving
wavelength
\begin{equation} 
q(t)=q(t_0)\frac{a(t)}{a(t_0)}
\end{equation}
Then the perturbations in density and proper velocity 
(scalar and vector perturbations) grow as
\begin{equation}
f_k(q(t),t) \approx e^{\int_{t_0}^t \lambda_k(q(t'),t') dt'}f_k(q(t_0),t_0)
\label{eq:Bonner-equation}
\end{equation}
where 
$\lambda_k(q,t)$ is the instantaneous growth
rate in the Jeans theory of wave-vector $q$ at time $t$,
and $k$ labels the linear modes.
Equation (\ref{eq:Bonner-equation}) does not agree completely with Lifshitz'
full solution, but is quite close in standard cosmological models. 
For an extended discussion, see \cite{ZeldovichNovikov}.

The linear modes can be classified into decaying modes, which are
fields of incompressible proper velocities, and growing modes,
which are linear combinations of density modes and potential
proper velocities.
If a mode actually grows or decays at time $t$ 
also depends on whether the
co-moving wave-length is smaller or greater than the 
instantaneous Jeans length.
On sufficiently large scales, the perturbations surviving the
linear regime are 
coupled  density and potential proper
velocity fluctuations.

The outcome of these considerations is that at length scales
much smaller then the radius of the Universe, but much larger than
the Jeans' length,
structure formation, as long as the solutions are single-stream,
is governed by the following system of equations: 

\begin{equation}
\label{s:system}
\left\{
\begin{array}{l}
\displaystyle{\partial_t\rho+3\frac{\dot{a}}{a}\rho +
\frac{1}{a}\overrightarrow{\nabla} \cdot  ({\rho}\overrightarrow{u})=0}
\\
\\
\displaystyle{\partial_t \overrightarrow{u}+
\frac{\dot{a}}{a}\overrightarrow{u}+
\frac{1}{a}(\overrightarrow{u} \cdot \overrightarrow{\nabla}) 
\overrightarrow{u}=
-\frac{1}{a} \overrightarrow{\nabla}\psi}
\\
\\ 
\displaystyle{\nabla^2\psi=4 \pi G a^2 (\rho-\rho_b)}  
\end{array}
\right.
\end{equation}
If we also assume that initial rotational proper
velocity fluctuations have been damped out in the linear decay, then
$\overrightarrow{u}$ is a potential field at some time $t_0$.
$\psi$ is 
the gravitational 
potential generated by the source $\rho-\rho_b$. We note that $\rho-\rho_b$ 
can be both negative and positive, and is zero on average.
On short time scales we can take $a$ constant, and by a change of scale
we can set it equal to one. On the kinetic level the system is
then described by the following Jeans-Vlasov-Poisson equation, valid also
after caustic formation:

\begin{equation}
\label{eq:system-JVP}
\left\{
\begin{array}{c}
\displaystyle{\partial_t f+\frac{\overrightarrow{p}}{m}\cdot
\overrightarrow{\nabla}_x f -\overrightarrow{\nabla}\psi\cdot
\overrightarrow{\nabla}_p f = 0}
\\
\\
\displaystyle{\rho = \int f\, d\,\overrightarrow{p}
\quad \overrightarrow{u}=\frac{1}{\rho}
\int \frac{\overrightarrow{p}}{m} f\,d\,\overrightarrow{p}
\quad \nabla^2\psi=4 \pi G (\rho-\rho_b)}  
\end{array}
\right.
\end{equation}
The initial conditions of equations (\ref{s:system})
and (\ref{eq:system-JVP}) can be taken to be the fluctuations at some 
stage of linear growth.
It is a remarkable fact that the fluctuations at one particular time during 
linear growth are in fact  observable in the fluctuations of the $3K$ 
blackbody background radiation 
\cite{Partridge,Page,SmootScott,KamionkowskiKosowsky}.
At red-shift $z \simeq 1000$ (age of the Universe $10^5$ years), photons    
fell out of thermal equilibrium with electrons and nuclei; what is observed 
today
as $3K$ black body radiation is the red-shifted spectrum of photons that were 
in equilibrium with matter at that time.
COBE observations measure the temperature of the blackbody radiation with a 
beam width of $7^{\rm o}$, and detect mean square variations of about $30 \mu K$ 
(one part in $10^5$). Laid out in the sky, COBE can thus be said to distinguish
a spherical grid of $50 \times 50$ patches, i.e. a decade and a half in each 
direction.
Experiments in the near future (MAP, Planck Surveyor) are expected to increase
the angular resolution of COBE by more than one order of magnitude. 
Over the range of COBE, observations are in agreement with the 
Harrison-Zeldovich\cite{Harrison1970,Zeldovich} prediction of Gaussian 
initial density
fluctuations
with spectrum \cite{SmootScott,Kogut}:  
\begin{equation}
\label{s:spectrum}
P(k)=A k^{n} \qquad n \simeq 1
\label{eq:harrison-zeldovich}
\end{equation}
At scales smaller than about $1^{\rm o}$, theoretical arguments predict
deviations from (\ref{eq:harrison-zeldovich}).
In cold dark matter-dominated models these are determined
by the CDM transfer function, which at intermediate scales,
in the present universe in the range
$10-50 h^{-1}\hbox{Mpc}$,
gives a plateau where fluctuations are also Gaussian
as in (\ref{eq:harrison-zeldovich}), but with
$n \simeq -1$ \cite{BardeenBondKaiserSzalay}.
We remark that in practically all cosmological
models, the spectrum 
(\ref{eq:harrison-zeldovich}) is not expected to be valid
in an arbitrary wide range, but to be modified at smaller scales.
For further recent discussions on the assumed limits of validity of
(\ref{eq:harrison-zeldovich}) and prospects of experimental
observations of such deviations, see e.g.
\cite{SmootScott,KamionkowskiKosowsky,MeiksinWhitePeackock}.

\section{The generalized Zeldovich solution: a Lagrangian integrable model}
\label{s:generalized-zeldovich}
We now turn to  
 a Lagrangian integration scheme where initial mass density 
$\rho$ is concentrated on a discrete set of particles. 
The algorithm which we are going to describe was invented in plasma
physics (i.e., equation
(\ref{eq:system-JVP}) with the opposite sign of $G$) in the early
'60ies, and already previously used in simulations of one-dimensional self-gravitating
systems \cite{BurganGutierrezMunierFijalkowFeix,Rouet1990,Rouet1991}.
The derivation we will give stresses
localized initial perturbations.  
Boundary conditions are hence those 
of an unperturbed quiescent state 
to the left and the right of the perturbation. As the perturbation
develops it will typically spread, and move into the 
quiescent state, thus inducing that to move.
The main ingredient in the derivation is a regularisation of
the differences between the (formally divergent) forces pulling the
particle to the left and to the right. 
Since the physical meaning of an infinite
classical self-gravitating system is that of an approximation 
(on sufficient small scales) to a system governed by general relativity,
and since the speed of propagation of the gravitational interaction
in general relativity is finite, localized
perturbations are relevant in this problem.

We now take space one-dimensional. 
The mass density distribution corresponding to a system of point-like 
particles is specified by: 
\begin{equation}
\rho(x,t)=\sum_{k} m_k\delta(x-x_k(t))
\label{initial}
\end{equation}
We require that the average density of the point-mass system is
the same as the background density, i.e.
\begin{equation}
\hbox{Lim}_{L\to\infty} \left[\frac{1}{2L}\sum_{k:x_k(t) 
\in [-L,L]}m_k \right] \, = \, \rho_0 \, = \,
\rho_b
\label{rho-average}
\end{equation}
and that the perturbation is localized
in a weak sense, such that 
as $L$ tends to infinity the measure of the point masses in
$[L,L+1]$ tends to the uniform measure in the same interval,
with convergence faster than $\frac{1}{L^2}$.
The one-dimensional gravitational potential is {\it formally} given by
\begin{equation}
\label{uau-formal}
\psi(x,t) = 2 G\int dy|x-y|(\rho(y,t)-\rho_b)\qquad\hbox{({\it formally})}
\nonumber
\end{equation}
With the initial conditions under consideration the integral
is convergent at infinity.
The regularisation referred to in the beginning of
this section is therefore to take
\begin{equation}
\label{uau}
\psi(x,t) = 2 G\, \hbox{Lim}_{L\to\infty}\left[
\sum_{l:{x}_l \in [-L,L]}m_l|x-x_l|
- \rho_b \frac{(x-L)^2 + (x+L)^2}{2}\right]
\end{equation}
where the limit exists by assumption.
As now $\psi$ is a known and well-defined function of
position, the equations of motion of the point masses are
\begin{equation}
\ddot{x}_k=-\frac{\partial \psi(x)}{\partial x}|_{x=x_k}
=-2 G \,\left[\hbox{Lim}_{L\to\infty}\sum_{l:{x}_l \in [-L,L]}m_l 
\hbox{sign}(x_k-x_l)- 2\rho_b x_k\right]
\label{illdef}
\end{equation}
with initial conditions 
$x_k|_{t=0}=x_k(0)$ and
$\dot{x}_k|_{t=0}=u_0(x_k(0))$.
Equation (\ref{illdef}) expresses that the force acting
on particle $k$ is equal to the difference of net mass, in excess of
the background, to the right and to the left of that particle.
Hence, were $L$ infinite,
it would formally be the sum of four terms (two with positive
and two with negative signs), each of which would be infinite. 
Our regularisation acts on the net force to the left and to
the right separately, by the requirement that the initial 
perturbation is localized. 

With periodic boundary conditions
our regularisation would not give a unique answer
since the limit (\ref{uau}) would not exist.
One could instead regularise 
separately the net force from
particles and net force from the background, for instance
by taking the net force of each kind initially to be 
zero, or assuming that the gravitational
force actually is of finite range, and then take the
limit when that range increases.
Using (\ref{illdef}) necessarily assumes
implicitly a regularisation, of which ours is a
physically transparent one, at the price that we consider only
sufficiently localized perturbations.

We now want to transform (\ref{illdef}) 
to the Eldridge-Feix scheme.
The acceleration acting on point mass $k$ at the initial time is
\begin{equation}
a(x_k(0),0)=-2G\,\left[ \hbox{Lim}_{L\to\infty}
\sum_{l:x_l \in [-L,L]}m_l \hbox{sign}(x_k(0)-
x_l(0))- 2\rho_b x_k(0)\right]
\label{eq:initial-forces}
\end{equation} 
From (\ref{illdef}) follows that as long as no other particle
overtakes particle $k$ we have the simple evolution law
\begin{equation}
\ddot{x}_k(t)=a(x_k(0),0)+4G \rho_b (x_k(t)-x_k(0))
\qquad \hbox{(time $t$ before collision)}
\label{eq:between-collisions}
\end{equation} 
When particle $l$ overtakes particle $k$, the gravitational force
acting between them changes sign.
We may therefore write
the equations of motion at an arbitrary time $t$ as
\begin{equation}
\label{newton}
\begin{array}{l}
\displaystyle{\ddot{x}_k(t) = a(x_k(0),0)+4G\rho_b\left(x_k(t)-x_k(0)\right)+}
\\
\\
\displaystyle{-2G \sum_{l} m_l \left[\hbox{sign}(x_k(t)-x_l(t))-
\hbox{sign}(x_k(0)-x_l(0))\right]}
\\
\\
\end{array}
\end{equation}
\\\\
If units are chosen such that $4G\rho_b=1$ (see below), the
mapping from one collision
to the next is then
\begin{equation}
\label{uauintegrated}
\begin{array}{l}
\displaystyle{x_k(t^{i+1}_{coll})=x_k(t^{i}_{coll})-a_k(t^i_{coll})+}
\\
\\
\displaystyle{+\frac{a_k(t^i_{coll})+
\dot{x}_k(t^i_{coll})}{2}\exp(t-t^{i}_{coll})+}
\\
\\
\displaystyle{+\frac{a_k(t^i_{coll})-
\dot{x}_k(t^i_{coll})}{2}\exp(t^{i}_{coll}-t)}  
\end{array}
\end{equation}
where $\dot{x}_k(t^i_{coll})$ and
$a_k(t^i_{coll})$ are the velocity and the acceleration
felt by the $k$'th particle at the 
preceding collision.

\section{Numerical experiments}
\label{s:numexp}
\begin{figure}[p] 
\vspace{-4mm}
 \epsfxsize=0.6\hsize
 \mbox{\hspace*{.18\hsize}
\epsffile{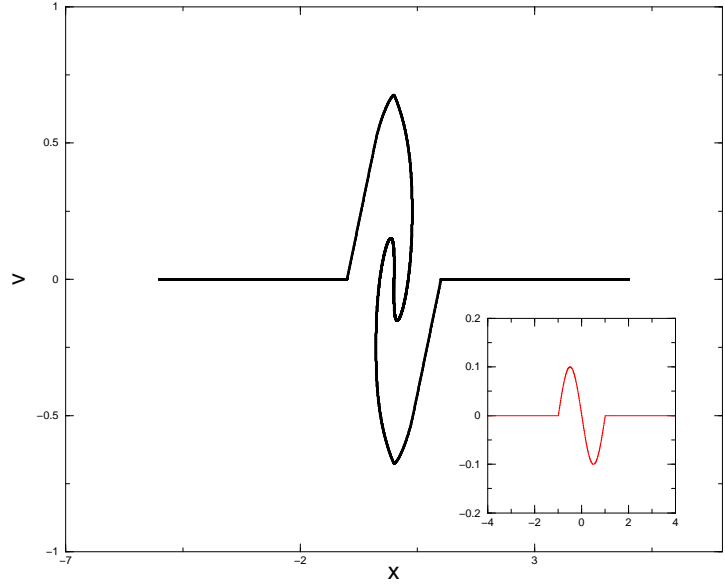}}
\vspace{0.5cm}
\caption[]
{Velocity field vs.  position at time  $t=4.0\, \omega_J^{-1}$.
Number of particles is $N=15 \times 10^{3}$, initial
sinusoidal conditions in lower right corner insert. 
}
\label{f:single}
\end{figure}

\begin{figure}[p] 
\vspace{-4mm}
\epsfxsize=0.6\hsize
\mbox{\hspace*{.18\hsize}
\epsffile{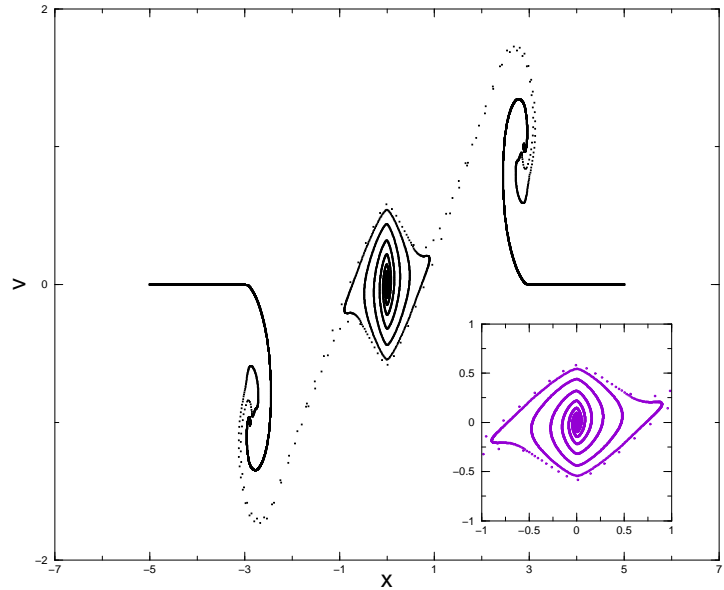}}
\vspace{0.5cm}
\caption[]
{Velocity field vs. position of the system in fig. \ref{f:single} 
at a later time, well into the non-linear regime 
($t=11.08\, \omega_J^{-1}$). The insert is a blow-up
of the central region. 
Three high-density peaks are clearly developed.}
\label{f:double}
\end{figure}

A characteristic scale in time is set by the Jeans' frequency,
related to the gravitational constant $G$ and the mean density
$\rho_0$ by
\begin{equation}
\omega_J = \sqrt{4G\rho_0}
\label{eq:Jeans-frequency}
\end{equation}
In the following we will measure time in units of 
$\omega_J^{-1}$. Since the mean density is equal
to the background density $\rho_b$, this means that
we choose a scale in time
such that the value
of the product $4G\rho_b$ 
in (\ref{newton}) is one.
\\\\
The model of self-gravitating particles without pressure is valid on spatial
scales much greater than the Jeans length $\lambda_{J}$. The intrinsic
spatial scale $\lambda_{J}$ is thus for us zero.
Characteristic spatial scales can hence only come from the initial
conditions. If the initial perturbation is ordered
(smooth function of spatial coordinate), and 
in analogy with (\ref{rho-average}) and (\ref{illdef}) has
support on an interval $[-L,L]$
then all scales can be measured in terms of $L$.
Temporal and spatial scales 
$\omega_J^{-1}$ and $L$ imply however a velocity scale $L\omega_J$,
and we can therefore separate ordered initial velocity perturbations 
as to if they are large or small compared to 
$L\omega_J$. 
An initial density perturbation $\delta \rho$
is
naturally measured in units of the mean density.
We choose to normalize mass in terms of the mass $M$ initially
involved in the perturbation, which means that $\rho_0$,
and therefore $\rho_b$, is $\frac{M}{2L}$.
The constant $G$ in
(\ref{newton}) is $\omega_J^2\frac{L}{2M}$, in our units 
hence $\frac{1}{2}$.
\\\\
The explicit formula (\ref{uauintegrated})  allows immediately for
a fast numerical scheme with operations count ${\cal O}(N)$ per collision,
where $N$ is the number of particles; this is scheme of
Eldridge and Feix\cite{EldridgeFeix}.
Elsewhere we will discuss an asymptotically more efficient version of 
the algorithm with an operation count of ${\cal O}(\ln N)$ per collision
\cite{AurellFanelliMuratoreNoullez}. 
We note that the computer time needed to advance the system up to time $t$
is proportional to the product of the number of operations per
collision and $N_{coll}(t)$, the number of collisions up to time $t$. 
If with initially smooth velocity and density perturbations, represented 
as $N$ discrete particles, the mean time between collisions depend on
$N$ as $\frac{1}{N}$, then in the algorithm used here
the operations count in advancing the system
up to intrinsic time $t$ would be $N^2$. 
However, due to stretching and separation, the discretization
becomes gradually a less accurate description of the smooth field to
be approximated; this effect is far from uniform in phase space,
and the putative ${\cal O}(N^2)$
operation count therefore only holds approximately
at an initial stage. 
\\\\
In fig~\ref{f:single} we show an initially smooth velocity perturbation
(insert) of sinusoidal shape in the interval $[-1,1]$. The initial
density perturbation is zero. The full system thus consists of a uniform
stationary background with density $\rho_b$, and a number of particles
as in~(\ref{initial}), distributed uniformly with respect to the same
density, and at rest outside the interval. As long as no particle from
the inside has reached the boundaries all three terms on the right-hand
side of equation~(\ref{newton}) remain unchanged for a particle outside
$[-1,1]$, and these particles thus remain at rest. The main figure
in fig~\ref{f:single} show the solution at time $4.0\,\omega_J^{-1}$,
at a resolution where obviously the continuum description is still
applicable, and where no particle from inside has reached the
boundaries. In phase space the distribution has support on a curve
of spiral shape. If initial velocity is large
or compared to $L\omega_J$,
the fastest particles typically reach the boundaries before a spiral
is formed, while if initial velocity is much smaller
than $L\omega_J$ one or more turns of the spiral
form before any particles reach the boundaries. These observations
have already been made by Rouet et al~\cite{Rouet1990,Rouet1991},
and show a qualitative difference in the solutions depending
on the initial kinetic energy.
\\\\
In fig~\ref{f:double} we show the development of the system in 
fig~\ref{f:single} at successive later stages. The most distinct
features are the formation of a high-density cluster in the middle,
and two expanding high-density fronts to the left and right,
but with very low density in between.
The two fronts form when particles from the interval collide with
particles that were initially at rest. The first particles to do so
overtake the same mass in resting particles and background, and will
therefore feel no change in the acceleration. The resting particles that
have been overtaken will on the other hand feel an acceleration towards
the front that has passed. They will hence start moving towards the
front, making non-leading particles in the front overtake 
more background than particles, and therefore also
feel an increased acceleration. Both these effects
lead to a high density concentration at the front, large force gradients,
and strong stretching in phase space. In fact, it is clear that with
the resolution used in  fig.~\ref{f:double},
at the fronts the continuum description is
already lost.
\\\\
In the center of the middle cluster the density is several times the
background, and the spirals are round, as in dynamics  
without the background, while in the outer parts, where
$\delta \rho$ is comparable to $\rho_b$, the spirals are deformed,
as above on fig~\ref{f:single}.  
\\\\
We now turn to  random conditions. We made three different choices of
velocity as function of spatial position: Brownian motion;
fractional Brownian 
motion with Hurst exponent equal to zero, and white noise.
All three are random Gaussian fields with power law spectra as in eq.
(\ref{s:spectrum}), of which we choose white noise and Brownian
motion because they are Markov processes, and have been investigated
in the context of the adhesion approximation~\cite{SheAurellFrisch,Sinai},
white noise also as a reference case, because it has earlier
been investigated by Rouet and collaborators~\cite{Rouet1990,Rouet1991},
and the fractional Brownian motion as an interesting
intermediate case.
\\\\ 
The translation between exponents in
(\ref{s:spectrum}) (density perturbations in 3D) and
our initial conditions (velocity perturbations in 1D)
is as follows: a Gaussian random function $f$ with stationary
increments has  second order
correlation function $<|f(x+l)-f(x)|^2>\sim l^{2h}$, where the Hurst
exponent $h\in[0,1[$ is related to the scaling exponent $n$ of the
spectrum $E(k)\sim k^n$,
through $2h+n+(D-1)=-1$. A scaling exponent $n_{1D}$ in $1D$
is therefore, for these purposes, analogous to a scaling exponent
$n_{3D}=n_{1D}-2$ in 3D. A density perturbation $\delta\rho_k$
is in the linear regime on the other hand tied to a velocity
perturbation $v_k \sim k\rho_k$. We therefore have $n^{(v)}=n^{(\rho)}+2$,
and combining the two relations $n^{(v)}_{1D}=n^{(\rho)}_{3D}$.
Our intermediate case with Hurst exponent equal to zero hence corresponds to
the Bardeen-Bond-Kaiser-Szalay intermediate spectrum 
with $n^{(\rho)}_{3D}=-1$, and will in
the following be referred to as BBKS initial conditions.
White noise and Brownian motion correspond to
$n^{(\rho)}_{3D}=0$ and $n^{(\rho)}_{3D}=-2$.


\begin{figure}[p] 
\vspace{-4mm}
 \epsfxsize=0.6\hsize
 \mbox{\hspace*{.18 \hsize}
\epsffile{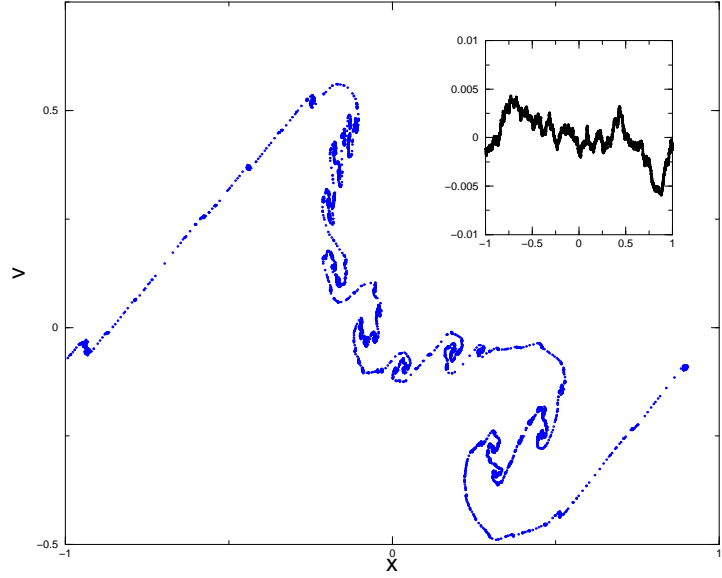}}
\vspace{0.5cm}
\caption[]
{Velocity field vs. position at time $4.76\, \omega_J^{-1}$,
starting from single-speed Brownian motion initial conditions
(small upper right insert). Number of simulated particles is $N=8192$.
}
\label{f:bm053}
\end{figure}

\begin{figure}[p] 
\vspace{-4mm}
 \epsfxsize=0.6\hsize
 \mbox{\hspace*{.18 \hsize}
\epsffile{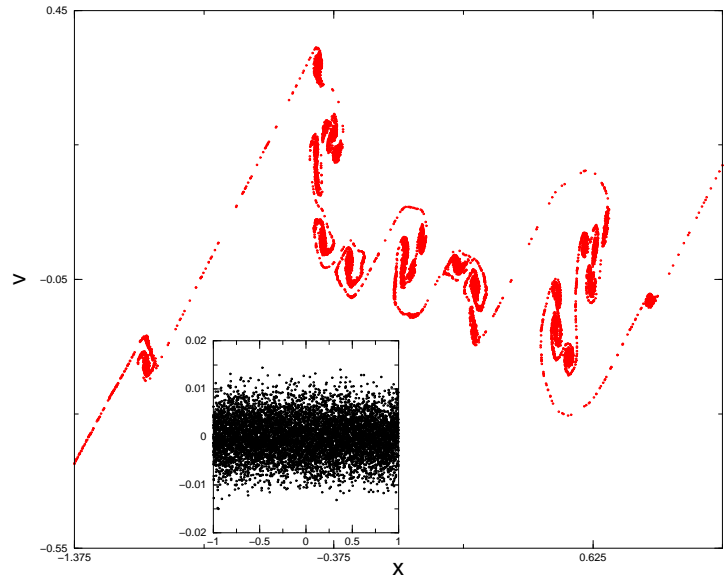}}
\vspace{0.5cm}
\caption[]
{Velocity field vs. position
at time $7.37\, \omega_J^{-1} $,
starting from single-speed white noise initial
conditions (lower insert).
The model is modified, as described in the main text,
to contain only background and no particles outside
the interval of the initial perturbation.
Number of simulated particles is $N=8192$.
}
\label{f:bm077}
\end{figure}

\begin{figure}[top] 
\vspace{-4mm}
 \epsfxsize=0.6\hsize
 \mbox{\hspace*{.18 \hsize}
\epsffile{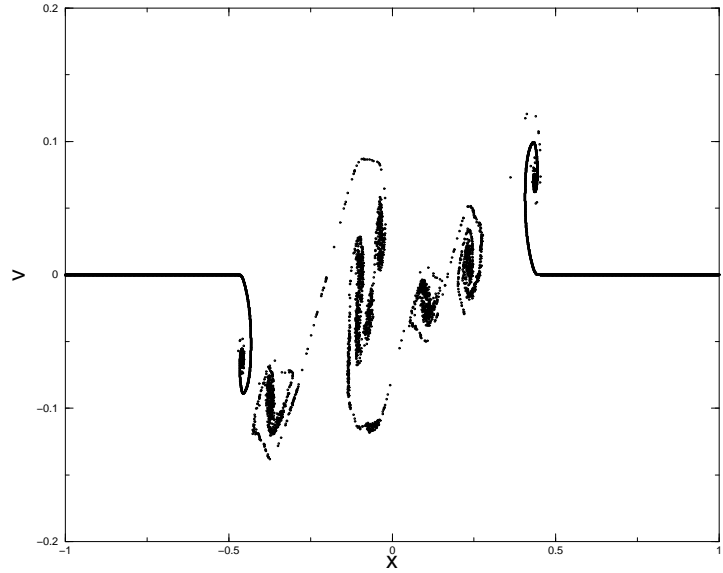}}
\vspace{0.5cm}
\caption[]
{Velocity field vs. position 
at time $7.37\, \omega_J^{-1} $,
starting from single-speed white noise 
initial conditions as in fig. \ref{f:bm077} inside
the interval of the perturbation, and quiescient particles
to the right and left thereof.
Note the formation of fronts as in fig. \ref{f:double}.
Number of particles initially involved in the perturbation,
with total mass one, is $N=8192$, and the total number of
simulated particles, including the ones to the left and right,
is $16384$.
}
\label{f:cl053}
\end{figure}

\begin{figure}[p] 
\vspace{-4mm}
 \epsfxsize=0.6\hsize
 \mbox{\hspace*{0.18 \hsize}
\epsffile{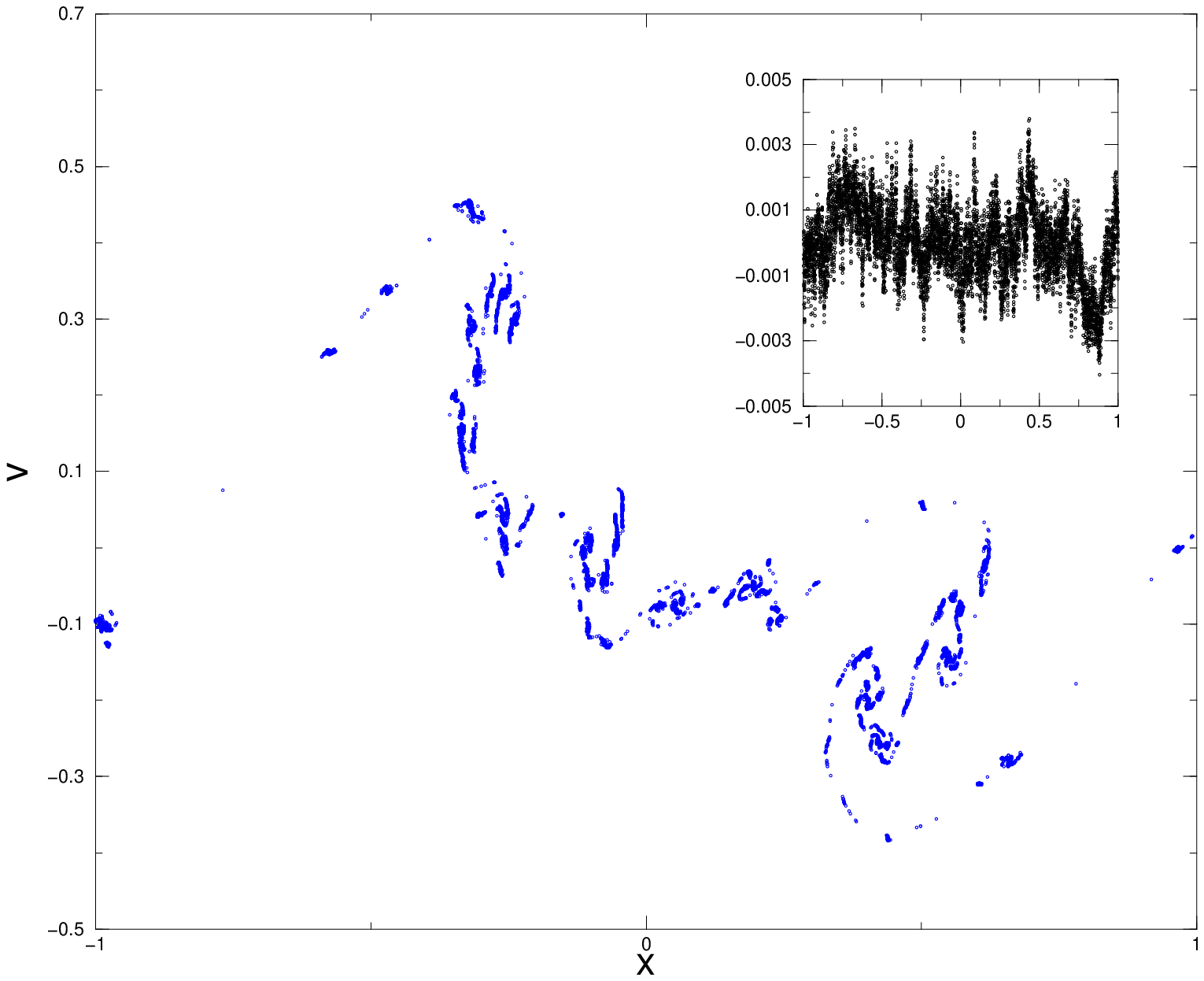}}
\vspace{0.5cm}
\caption[]
{Velocity field vs. position 
at time $t=6.7\, \omega_J^{-1}$
starting from single-speed BBKS initial conditions
(upper right insert).
Number of simulated particles $N= 8192$.
}
\label{f:z011}
\end{figure}

Gaussian random
fields are conveniently generated in the Fourier space 
representation.
At time $t=0$ particles are
uniformly distributed on an interval of 
size $\ell$ with unit spacing $\Delta x=\ell/N$  
and velocity 
\begin{equation}
v(x)=\Sigma_{k} v_k e^{ikx}
\label{randomfield}
\end{equation}
The sum over $k$ extends from $- \pi\,/\,\Delta x$ to 
$\pi \, /\,\Delta x$ in steps of $2 \,\pi \,/\,\ell$ and $v_{-k}=v_{k}^{*}$.
The Fourier components of positive $k$ are then chosen as a random Gaussian 
independent variables with variances:
\begin{equation}
<|v_k|^2>=\frac{\,\,k^{-2h-1}}{2\,\Delta x}
\label{variance}
\end{equation}
The field generated by (\ref{randomfield}) and (\ref{variance})
will be periodic with period $\ell$.
If $h$ is in the interval 
$[0,1[$ the field will have stationary increments on length
scales much less than $\ell$, and on these scales
approximate  a Fractional Brownian motion with Hurst exponent
$h$, while if $h$ is in the interval 
$]-1,0[$ the field itself will be stationary
on length scales much less than $\ell$, and approximate the derivative
of a Fractional Brownian motion. We choose $\ell = 2L$, so that particles 
are initially distributed in a box $[-L,L]$.
\\\\
We will now discuss units of time and space with random initial
conditions. One scale is $2 L$, another is the ultra-violet cut-off
$l_{UV}$, which in our case is at least as large as the initial
particle distance $\Delta x$, and a third is an infra-red cut-off
$l_{IR}$, which is not larger than $2 L$.
With $h$ in the interval 
$[0,1[$ we have in law
\begin{equation}
v(x+l)-v(x)\sim v_{UV}(\frac{l}{l_{UV}})^{h}
\label{eq:Hurst}
\end{equation}
where $v_{UV}$ is the size of typical velocity
fluctuations on the ultra-violet cut-off scale.
The typical overturning time at scale $l$ is then
\begin{equation}
t_l \sim \frac{l_{UV}}{v_{UV}}(\frac{l}{l_{UV}})^{1-h}  \qquad h\in[0,1[
\label{eq:turnover}
\end{equation}
The time 
$\frac{l_{UV}}{v_{UV}}$ can be measured in units of
inverse Jeans' frequency, and be small or large in those
units. Equation~(\ref{eq:turnover}) then predicts that
the characteristic time to form a structure of size
$l$ increases with $l$, such that small scales form first.
At sufficiently large scales $l$ the initial 
velocity fluctuations
will be small compared to $l\omega_J$, and we hence 
expect to see the central spiral structure of figs~\ref{f:single}
and~\ref{f:double}, but not much of the fronts.
\\\\
On the other hand, if the spectral exponent is larger than $-1$
(Hurst exponent less than zero), then the initial velocity
field is homogeneous, characterized by a rms velocity
$v_{RMS}$, which is on the order of $v_{UV}$, and
we therefore expect the simpler result
$t_l \sim \frac{l_{UV}}{v_{RMS}}(\frac{l}{l_{UV}})$.
At scales much larger than $l_{UV}$
velocity fluctuations are again much smaller than 
$l\omega_J$, and we therefore
expect mainly to see the central
spiral structures of fig~\ref{f:single}
and fig~\ref{f:double}.
We remark that if we reason by analogy, and assimilate
these structures to shock waves in the adhesion
model, which trap fast particles, then the characteristic
times will be longer, and in fact again of the
form~(\ref{eq:turnover}). The turn-over
time $t_l$ then grows faster then linearly
with 
$l$~\cite{GurbatovSimdyankinAurellFrischToth,GurbatovSaichevShandarin}.
Unfortunately our present resolution is not sufficient for
a precise determination 
the typical size as function of time and
of the temporal development of the
spectral shape of the perturbations.
\\\\
In figs.~\ref{f:bm053}, \ref{f:bm077}, 
\ref{f:cl053},
\ref{f:z011} we show the phase space portraits 
with Brownian motion, white noise and BBKS initial conditions,
respectively. As expected, we see many spiral structures, small
and large, and fronts at the left and right
boundaries, except in fig.~\ref{f:bm077}, included as
a consistency check (see below). 
In qualitative agreement with the adhesion
model, we also see ``ramps'' with low density, and where
velocity is an increasing function of position, interspersed
with regions of high density. 
By visual inspection of an agglomeration of a certain scale,
it appears that the velocity to the right of such an agglomeration
is less than on the left, such that velocity has negative gradient
through the agglomeration, just as in shocks
in the adhesion model.
This picture is however complicated by the fact that there are
agglomerations of different sizes, and that they are not simply
ordered from left to right. A certain stretch of phase space,
say just to the left of some structure, is on some other scale
included inside a much larger structure; this effect appears
to be especially pronounced with Brownian motion initial
conditions.
In the adhesion model such microstructures are of course
hidden inside the shocks.
\\\\
The difference between figs. \ref{f:bm077}
and~\ref{f:cl053} is that in fig. \ref{f:cl053}
we include, following our general approach,
a medium of quiescient particles to the right and
left. As in fig.~\ref{f:double} we then find two expanding fronts,
where the particles of initial velocities around $v_{RMS}$ in the
interval collide with the particles initially at rest. 
To show that the fronts do not qualitatively change 
the dynamics in the central region, 
we show for comparison
in fig.~\ref{f:bm077} a system
with only background outside the interval.
Particles that escape the central interval are then
accelerated outwards by the anti-harmonic force
in~(\ref{newton}), and give phase space portraits
where velocity $v$ is an increasing function of
position $x$ outside the interval.
\\\\
Another way to eliminate the fronts would be to take a perturbation
which is similar to white noise, but of which the envelope
would go smoothly to zero outside an interval of length $2L$,
which can be achieved with essentially flat spectrum at
wave numbers larger than $\frac{1}{2L}$. 
For Brownian and BBKS initial conditions we do not have to
worry about the fronts. Without changing any many-point statistics
we can constrain a realization using~(\ref{randomfield})
and~(\ref{variance}) to vanish
at one point, say $x=-L$, and then by periodicity also at
$x=L$. By~(\ref{eq:Hurst}) the typical velocities in the center
of the interval $[-L,L]$ will then be much larger
than at the boundaries, and the most prominent structures are
found there.


\begin{figure}[p] 
\vspace{-4mm}
\epsfxsize=0.6\hsize
\mbox{\hspace*{.18 \hsize}
\epsffile{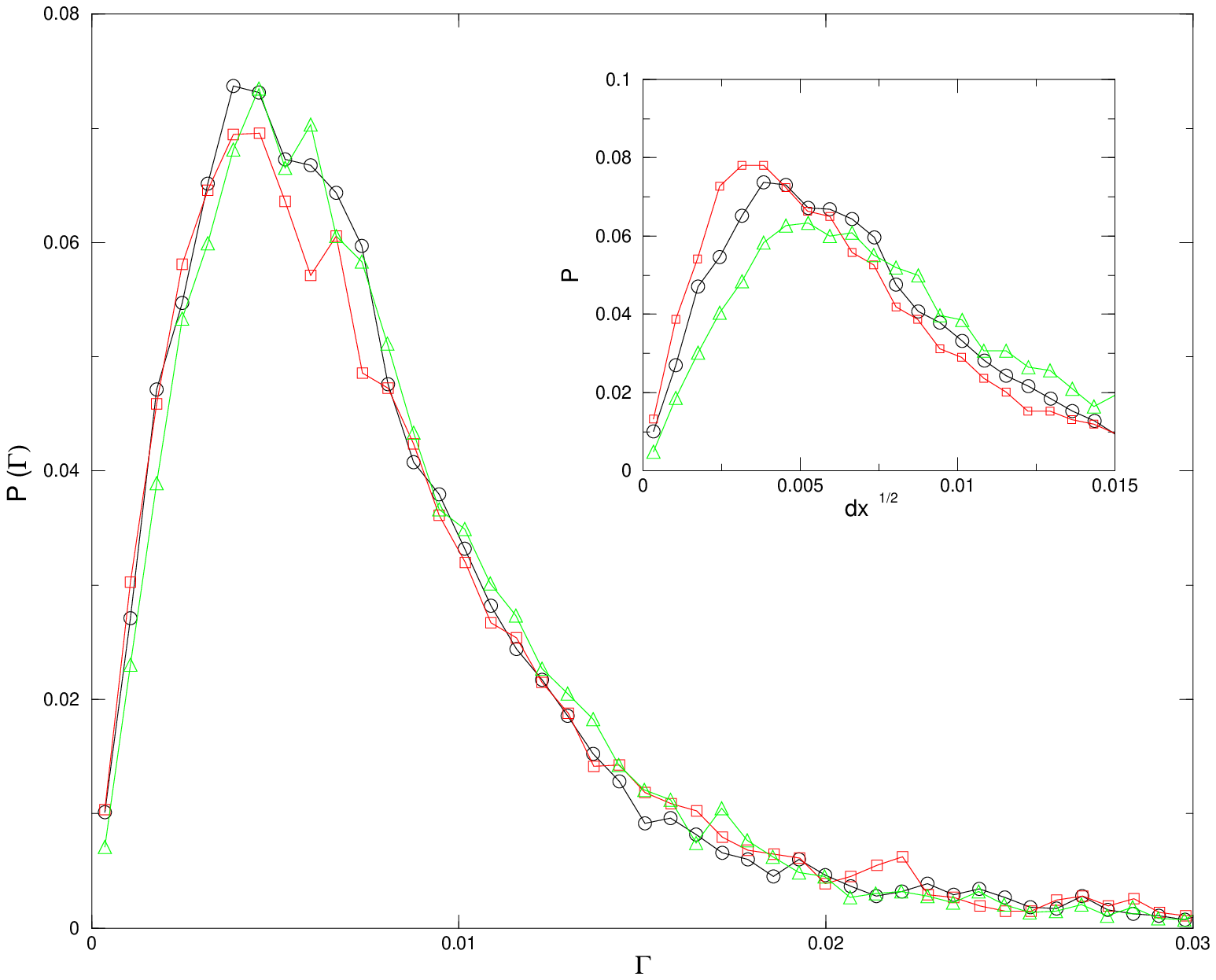}}
\vspace{0.5cm}
\caption[]
{
 Insert on the right: the PDF of the inter-particle distances to 
the power $1/2$, for Brownian 
initial velocities (squares, $t=6.4\omega_J^{-1}$), 
BBKS (circles, $t=7.5\omega_J^{-1}$) and 
white noise (triangles up, $t=8.1\omega_J^{-1}$).
Main figure: PDF of $\Gamma$, where 
$\Gamma=(\frac{dx}{\beta^h})^{1/2}$ and $\beta=0.6$.
}
\label{f:llbm}
\end{figure}

To quantify how accurate is a finite-particle description we 
looked at PDF of the 
inter-particle distances at different times. 
In fig.~\ref{f:llbm} we show the PDF of the inter-particle distances 
to the power one-half (insert on the right). 
In each case a peak is clearly observed:
the positions of these peaks are independent 
of the initial realizations and 
evolution time.
The main plot in fig.~\ref{f:llbm} shows the PDF of the rescaled quantity 
$\Gamma=(\frac{dx}{\beta^h})^{1/2}$ where 
the constant $\beta=0.6$ was obtained by a numerical fit.
We have no good explanation of the observed
scaling behaviour
at this point.

\section{Mass analysis}
\label{s:mass}

Qualitatively, the velocity field recalls the results obtained in the 
framework of the adhesion model 
\cite{GurbatovSaichevShandarin,SheAurellFrisch,VergassolaDubrulleFrischNoullez}, 
where ``ramps'' appear that separate shock regions where fully inelastic 
collisions among the Lagrangian particles occur.
It is interesting to try to introduce quantitative 
characteristics of the mass distribution in order to 
make a closer comparison.

\begin{figure}[p]
\vspace{-4mm}
\epsfxsize=0.6\hsize
\mbox{\hspace*{.18 \hsize}
\epsffile{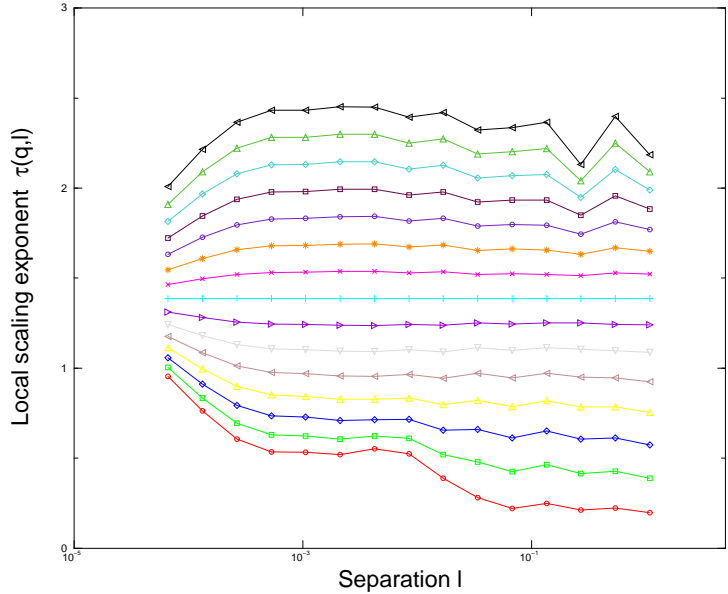}}
\vspace{0.5cm}
\caption[]
{
Local scaling exponent obtained as the logarithmic derivative of $M(q,l)$ 
versus $l$ at time
 $t=5.8\, \omega_J^{-1}$ with Brownian
initial velocities.
Number of particles is $N=8192$. The exponent $q$ ranges from
$0.125$ to $1.875$ in steps of $0.125$. 
}
\label{f:multi}
\end{figure}

\begin{figure}[p]
\vspace{-4mm}
 \epsfxsize=0.6\hsize
 \mbox{\hspace*{.18 \hsize}
\epsffile{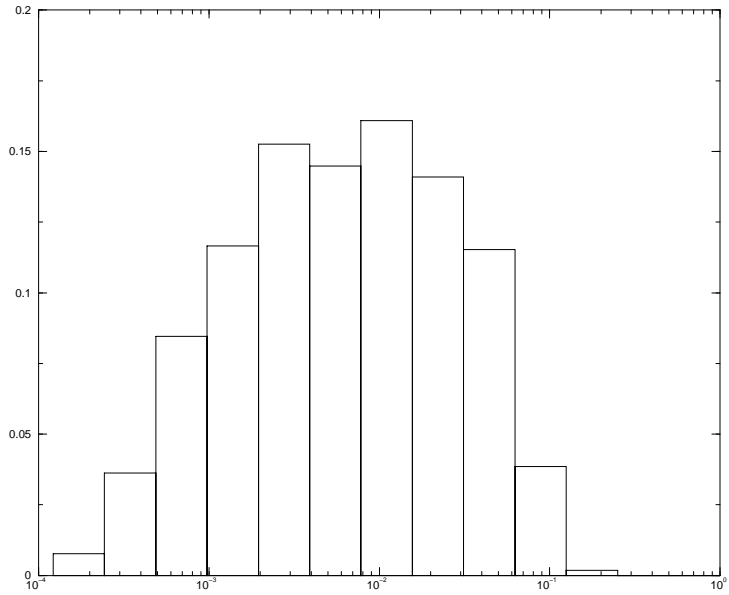}}
\vspace{0.5cm}
\caption[]
{
Normalized mass distribution in octaves for Brownian initial velocities
at $t=6.0\, \omega_J^{-1}$.
The number of particles is $8192$, number of independent
realizations $30$. Each realization was further averaged over
$10^4$ collisions, about $2 \times 10^{-3}\omega_J^{-1}$ in intrinsic time.
The height of a column
is the fraction of a total number of bins
containing mass in the range $[m,2m]$.
The bin size is $l=0.003125$, a
uniform density would hence correspond to a single column
at abscissa  $0.003125$ and ordinate $1.0$.

}
\label{f:mssbm}
\end{figure}

\begin{figure}[p!] 
\vspace{-4mm}
 \epsfxsize=0.6\hsize
 \mbox{\hspace*{0.18 \hsize}
\epsffile{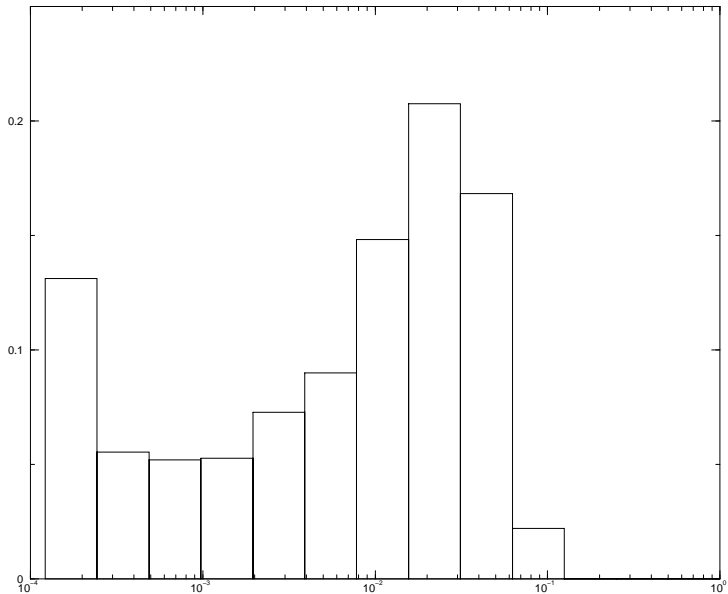}}
\vspace{0.5cm}
\caption[]
{Normalized mass distribution in octaves for BBKS initial velocities
at $t=5.9\, \omega_J^{-1}$,
Number of particles and bin sizes as in fig. \ref{f:mssbm},
number of independent realizations $20$.
}
\label{f:mssz}
\end{figure}

\begin{figure}[p!]
\vspace{-4mm}
 \epsfxsize=0.6\hsize
 \mbox{\hspace*{0.18 \hsize}
\epsffile{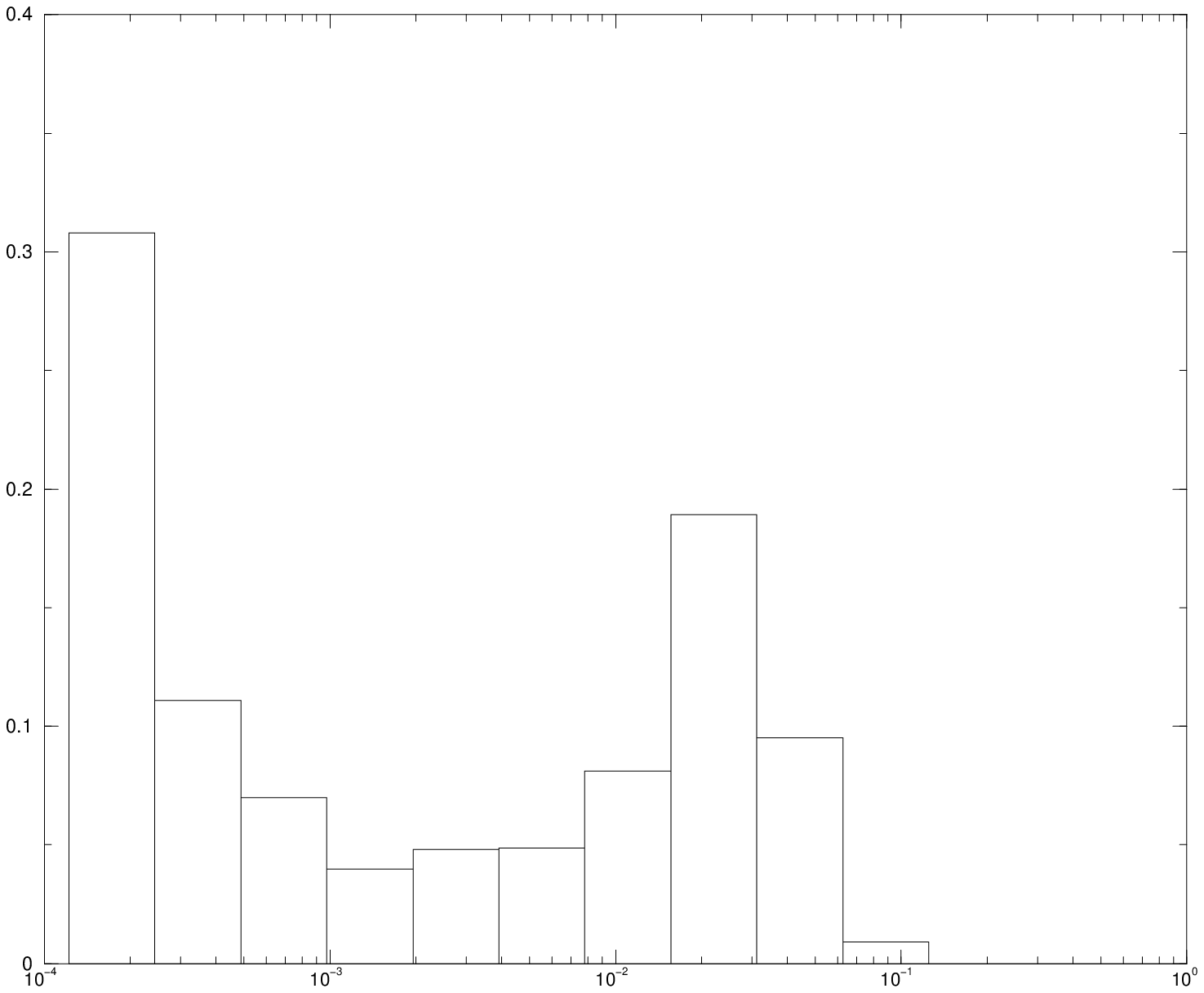}}
\vspace{0.5cm}
\caption[]
{
Normalized mass distribution in octaves for white noise initial velocities
at $t=7.0\, \omega_J^{-1}$.
Number of particles and bin sizes as in
fig. \ref{f:mssbm}, number of independent realizations $10$.
}
\label{f:msswn}
\end{figure}

One main result of 
\cite{SheAurellFrisch} and~\cite{AurellFrischNoullezBlank},
is that for $h\in[0,1[$ 
the inverse Lagrangian map, initial versus final
position, has a bifractal structure, similar to that of the Devil's
staircase construction using the standard triadic Cantor set.
Except for a set of
measure zero, all Lagrangian initial points land on shocks,
which in Eulerian coordinates are at definite shock locations,
and where therefore all of the mass is concentrated. 
The number density per unit length
of shock locations holding mass $m$ is in a well-defined
range distributed as a power-law $m^{-h-1}$. Most mass therefore
lies in the largest shocks formed at a given time, but the
number of smaller shocks is divergent. For the Brownian
motion initial condition it can be proven, and for the
other cases it is conjectured, that
Eulerian shock locations are almost surely dense. Most
of these shocks are still small.
However, since all mass
initially uniformly distributed after an arbitrarily short
time falls into a shock, the mass 
contained in an interval
after a finite time is in this model
only made up of the mass in the shocks actually inside
the interval.
The mass measure of the adhesion model with these initial
conditions is therefore bifractal, which can
be quantified by the scaling exponents of its moments 
\begin{equation}
M(q,l)= \Sigma_{i=1}^{B}\,p(x_i)^q\,l \sim l^{\tau(q)}
\label{moment}
\end{equation}
where $l$ is the length of the coarse-graining mesh, $B$ is the number
of boxes in the mesh, and $L=B l$.
The sum is normalized such that $\tau(1)=1$ 
and $\tau(0)=0$. 
At sufficiently
large $q$, where the threshold lies at 
$h$, the sum in (\ref{moment}) is dominated by a small number of
terms, corresponding to the intervals containing the larger shocks.
The exponent  $\tau$ in this range is then one.
At small $q$ (\ref{moment}) would instead be dominated by
almost empty intervals, each of which carries 
a mass $l^{\frac{1}{h}}$,
and of which there would be $l^{-1}$ in number. The scaling
exponent in this range would hence be $\tau(q)=\frac{q}{h}$.
With $n$ in the range $[-1,1]$ (which formally 
corresponds to $h$ in the range
$[-1,0]$), of which one case is white noise 
($n=0$, $h=-\frac{1}{2}$)
all the above statements remain true in the adhesion model,
but somewhat trivial. 
Shock locations holding
mass $m$ are still distrubuted as $m^{-h-1}$, but since
$h$ is now negative, most shocks are within an octave in size
of the largest. Shocks are not dense, the mass distribution
is almost surely concentrated on a finite number of points
per unit length, and $\tau(q)=1$ for all positive $q$.
\\\\
If the bifractal scaling behaviour of
the moments in (\ref{moment}) would be observed
also for a self-gravitating system, then the two models
would in this sense be equivalent.
Possible deviations from bifractality
would on the one hand be intrinsic effects of the self-gravitating
dynamics.
In fig.~\ref{f:multi} the local scaling exponent $\tau=\tau(q,l)$ is 
shown as a function of $l$.
One problem, discussed at length in
\cite{AurellFrischNoullezBlank} for the adhesion model,
is that with a finite number of particles, true scaling behaviour
can only be observed in a range where most intervals
actually contain a particle. 
At smaller mesh sizes, most boxes will be empty, and
(\ref{moment}) would be dominated by a small fraction
at any positive value of $q$, which gives the spurious result
$\tau(q)=1$, also for $q$ in the interval
$[0,h]$. The predicted cut-off occurs at $l_{sp}\sim \epsilon^{h}$,
where $\epsilon$ is the numerical mesh size in a simulation of
Burgers' equation, and which we in analogy in our case can
take to be $2L/N$. The value of $l_{sp}$ is thus
unfortunately not very small.
In the present problem mass is not so concentrated, and
we could expect to perhaps see a slightly wider range.
Nevertheless, it is clear that in the range $q$ less than one,
the only cleanly observed scaling behaviour in $l$ is the spurious one
at small $l$, in our case $l$ less than about $10^{-4}$.
\\\\
A more direct quantification
of the mass distribution
is the mass octave function (MOF), 
i.e. the probability to 
find a non zero contribution to the mass density as function of the mass 
itself, coarse-grained in octaves.
As discussed above, in 
the adhesion model the number of intervals with mass in an octave
around $m$ scales as 
$m^{-h}$~\cite{SheAurellFrisch,VergassolaDubrulleFrischNoullez},
and this behaviour would be observable in all three models
we study.
\\\\
At a less detailed level, 
the MOF measures the degree of dishomogeneity of a probability distribution,
since  in
terms of the MOF a uniform distribution corresponds to a logarithmic 
histogram with only one non-zero entry. 
\\\\
Figs.~\ref{f:mssbm}, \ref{f:mssz} and~\ref{f:msswn}
show the MOF for Brownian motion, BBKS and white noise
initial conditions.
As we see, the observation of a non-uniform density distribution
is clearly borne out, since the MOF diagrams have support
over several octaves in all three cases.
More interestingly, although statistics and the range are
not very large, for white noise initial conditions the
most massive boxes are the most frequent ones, while 
for Brownian motion less massive boxes are
at least equally frequent, in qualitative agreement
with the adhesion model. The BBKS model seems to be intermediate.


\section{Discussion}
\label{s:discussion}
Observed structures in our universe  
are believed to have been caused by gravitational instabilities
in an initially almost homogeneous medium.
The temperature fluctuations of the Cosmic Microwave Background
radiation give direct observational access on the primordial 
density and proper velocity fluctuations which seeded
these structures, albeit as of today
only on a limited range of scales.
\\\\
The general problem of structure formation by nonlinear deterministic evolution
equations, acting on random initial conditions, is a topic of much
current interest, particularly in the context of Burgers' turbulence,
where progress has been made using the special integrability
properties of Burgers' equation.
In this paper we have investigated the dynamics
of a one-dimensional
self-gravitating medium, as a more accurate
model of structure formation in the universe. Therefore, it is of interest
to understand how different are the solutions to Burgers' equation and
self-gravitating systems starting from the same initial data.
\\\\
The problem addressed is hence
not to determine if these two models are
closely similar in general, but if they are
similar with specific initial conditions
suggested by cosmology. We have focused on Gaussian random fields
with scaling spectrum, of the type originally suggested by
Harrison and Zeldovich, but with a variable scaling exponent.
The main obstacle to a detailed comparison is then to
solve numerically the self-gravitating system, since 
Burgers' equation is solved directly with
the Hopf-Cole transformation.
In one spatial dimension we have been able to use
an efficient numerical 
algorithm
which exploits a special Lagrangian
quasi-invariance of the gravitational force between particle
collisions.
\\\\
A self-consistent formulation of an infinite
self-gravitating system demands
a background term from the average mass density.
This somewhat trivial term
changes the properties of the solutions qualitatively
compared to those of a finite mass concentration,
with zero background.
A finite self-gravitating system has
a definite center of mass, and particles which are
furthest from the center feel the strongest attraction,
while in the self-gravitating system with background
mass far from the initial perturbation feel no
attraction at all. As soon as the system with background 
has developed structures of much higher density than
the background their further development is however
quite similar to a finite self-gravitating system:
this shows up in the formation
of a central body of enhanced density and spiral
phase-space structures.
A mathematical difference between perturbations in a finite
and an infinite self-gravitating system is an effective
anti-harmonic term in the latter, which appears when 
particle density thins out. Since if in the infinite
self-gravitating system density can only be less than the
average locally if it is higher elsewhere, a thinning out
assumes an attractive agglomeration at some other nearby
position, which can be seen as
the cause of that extra force.
\\\\
One prediction with analogy with the adhesion model is
that for initial perturbations with strong support at low
wave numbers, as our case Brownian motion, we expect
to see mass agglomerations of very different sizes,
while for white noise initial conditions we expect to
find most mass agglomerations of similar size.
Figs.~\ref{f:mssbm}, \ref{f:msswn}
indeed show this behaviour, while
fig.~\ref{f:mssz} is intermediate.
\\\\
Summing up, we have shown that one-dimensional self-gravitating
dynamics can be investigated quantitatively in systems with
a large scaling range in the initial conditions.
Nevertheless, the problem remains computationally more demanding
than e.g. Burgers' equation, and our results
on the fractal properties of the mass distribution are
not conclusive.
An improved numerical procedure has been proposed by
A.~Noullez~\cite{Noullezprivate}, which would allow us to
significantly enhance the resolution and the statistics.
Results of this second stage of the investigation will
be presented in a forthcoming separate 
contribution~\cite{AurellFanelliMuratoreNoullez}.

\section*{Acknowledgments}
\label{s:acknowledgements}
We thank two anonymous referees for very relevant remarks,
in particular referee~A for pointing out a mistake in our earlier
numerical simulations, and that the algorithm we had constructed 
was in fact the same as that of 
Eldridge and Feix~\cite{EldridgeFeix}.
We thank referee~B for help with the cosmological background,
and for pointing out reference~\cite{BardeenBondKaiserSzalay}.
We thank K.H. Andersen, S.N.Gurbatov, U. Frisch,
M.van Hecke and
A. Noullez
for discussions.
This work was supported by RFBR-INTAS 95-IN-RU-0723 (E.A. and D.F.), 
by the Swedish Natural Science
Research Council through grants M-AA/FU/MA~01778-333 (E.A.)
and M-AA/FU/MA~01778-334 (D.F), and by European Community
Human Capital and Mobility Grant ERB4001GT962476 (P.M.G.).
E.A. and P.M.G thank the European Science Foundation and the
local organizers of the TAO programme Study Center (Mallorca, 1999)
for an opportunity to write up this work.

\end{document}